\documentclass[a4paper,12pt]{article}
\usepackage {amsmath}
\usepackage {amsxtra}
\usepackage {amsbsy}
\usepackage {amscd}
\usepackage {latexsym}
\usepackage {amssymb}
\usepackage {amsfonts}
\usepackage {a4wide}
\usepackage {indentfirst}
\usepackage {cite}
\usepackage {bm}

\newcommand {\oks}[2]{{\raise0.7ex\hbox{${\scriptstyle #1}$}\!\mathord{\left/
{\vphantom{{1}{2}}}\right.\kern-\nulldelimiterspace}\!\lower0.7ex
\hbox{${\scriptstyle #2}$}}}

\begin{document}

\title{Generalized Dirac-Pauli equation and
neutrino quantum states in matter}
\author{Alexander Studenikin{\thanks{
Address: Department of Theoretical Physics, Moscow State
University, 119992  Moscow, Russia,
 e-mail:studenik@srd.sinp.msu.ru .}},
Alexei Ternov{\thanks{Department of Theoretical Physics, Moscow
Institute for Physics and Technology, 141700 Dolgoprudny, Russia,
e-mail:A\_Ternov@mail.ru }}}
  \date{}
  \maketitle

\begin{abstract}{Starting with the Dirac-Pauli equation
for a massive neutrino in an external magnetic field, we propose a
new quantum equation for a neutrino in the presence of background
matter. On this basis the quantum theory of a neutrino moving in
the background matter is developed: i) for the particular case of
matter with constant density the exact solutions of this new
equation are found and classified over the neutrino spin states,
ii)  the corresponding energy spectrum is also derived accounting
for the neutrino helicity. Using these solutions we develop the
quantum theory of the spin light of neutrino ($SL\nu$) in matter.
The $SL\nu$ radiation rate and total power are derived for
different linear and circular polarizations of the emitted
photons. Within the solid base of the developed quantum approach,
the existence of the neutrino self-polarization effect in matter
is also shown.}

\end{abstract}

Recently in a series of our papers
\cite{EgoLobStu00LobStu01DvoStuJHEP02StuPAN04} we have developed
the quasi-classical approach to the massive neutrino spin
evolution in the presence of external electromagnetic fields and
background matter. In particular, we have shown that the well
known Bargmann-Michel-Telegdi (BMT) equation \cite{BarMicTelPRL59}
of the electrodynamics can be generalized for the case of a
neutrino moving in the background matter and under the influence
of external electromagnetic fields. The proposed new equation for
a neutrino, which simultaneously accounts  for the electromagnetic
interaction with external fields and also for the weak interaction
with particles of the background matter, was obtained from the BMT
equation by the following substitution of the electromagnetic
field tensor $F_{\mu\nu}=(\bf E,\bf B)$:
\begin{equation}\label{sub}
F_{\mu\nu} \rightarrow E_{\mu\nu}= F_{\mu\nu}+G_{\mu\nu},
\end{equation}
where the tensor $G_{\mu\nu}=(-{\bf P},{\bf M})$ accounts for the
neutrino interactions with particles of the environment. The
substitution (\ref{sub}) implies that in the presence of matter
the magnetic $\bf B$ and electric $\bf E$ fields are shifted by
the vectors $\bf M$ and $\bf P$, respectively:
\begin{equation}
\bf B \rightarrow \bf B +\bf M, \ \ \bf E \rightarrow \bf E - \bf
P. \label{11}
\end{equation}
We have also shown  how to construct the tensor $G_{\mu \nu}$ with
the use of the neutrino speed, matter speed, and matter
polarization four-vectors.

Within the developed quasi-classical approach to the neutrino spin
evolution we have also considered
\cite{LobStuPLB03,DvoGriStuIJMP04,LobStuPLB04} a new type of
electromagnetic radiation by a neutrino moving in the background
matter and/or gravitational fields which we have named the "spin
light of neutrino" ($ SL\nu$). The $ SL\nu$ originates, however,
from the quantum spin flip transitions and for sure it is
important to revise the calculations of the rate and total power
of the $ SL\nu$ in matter using the quantum theory. Note that
within the quantum theory the radiation emitted by a neutrino
moving in a magnetic field was also considered  in
\cite{BorZhuTer88}.

In this paper we should like to present a reasonable step forward,
which we have made recently, in the study of the neutrino
interaction in the background matter and external fields. On the
basis of the generalization of the Dirac-Pauli equation of the
quantum electrodynamics we propose a new quantum equation for the
neutrino wave function with effects of the neutrino-matter
interaction being accounted for. This new equation establishes the
basis for the quantum treatment of a neutrino moving in the
presence of the background matter. In the limit of the constant
matter density, we get the exact solutions of this equation,
classify them over the neutrino helicity states and determine the
energy spectrum, which depends on the helicity. Then with the use
of these wave functions we develop the quantum theory of the $
SL\nu$ and calculate the rate and power of the spin-light
radiation in matter accounting for the emitted photons
polarization. The existence of the neutrino-spin self-polarization
effect \cite{LobStuPLB03,LobStuPLB04} in matter is also confirmed
within the developed quantum approach \footnote{The neutrino-spin
self-polarization effect in the magnetic field was discussed in
\cite {BorZhuTer88}.}.

To derive the quantum equation for the neutrino wave function in
the background matter we start with the well-known Dirac-Pauli
equation for a neutral fermion with non-zero magnetic moment. For
a massive neutrino moving in an electromagnetic field $F_{\mu
\nu}$ this equation is given by
\begin{equation}\label{D_P}
\Big( i\gamma^{\mu}\partial_{\mu} - m -\frac{\mu}{2}\sigma ^{\mu
\nu}F_{\mu \nu}\Big)\Psi(x)=0,
\end{equation}
where $m$ and $\mu$ are the neutrino mass and magnetic
moment\footnote{For the recent studies of a massive neutrino
electromagnetic properties, including discussion on the neutrino
magnetic moment, see Ref.\cite{DvoStuPRD_JETP04}}, $\sigma^{\mu
\nu}=i/2 \big(\gamma^{\mu }\gamma^{\nu}-\gamma^{\nu}
\gamma^{\mu}\big)$. It worth to be noted here that Eq.({\ref{D_P})
can be obtained in the linear approximation over the
electromagnetic field from the Dirac-Schwinger equation, which in
the case of the neutrino takes the following form
\cite{BorZhuTer88}:
\begin{equation}\label{D_S}
(i\gamma^\mu\partial_\mu-m) \Psi (x) =\ \int M^{F}(x',x)\Psi
(x')dx' ,
\end{equation}
where  $M^{F}(x',x)$ is the neutrino mass operator in the presence
of the external electromagnetic field.

For the case of the external magnetic filed, the Hamiltonian form
of the equation (\ref{D_P}) reads
\begin{equation}\label{H_form}
i\frac{\partial}{\partial t}\Psi({\bf r},t)=\hat H_{F}\Psi({\bf
r},t),
\end{equation}
where
\begin{equation}\label{H}
  \hat H_{F}=\hat {\bm{\alpha}} {\bf p} + \hat {\beta}m + \hat V_F,
  \hat V_F= -\mu \hat {\beta} {\bf {\hat \Sigma}} {\bf B},
\end{equation}
and ${\bf B}$ is the magnetic field vector. We use the Pauli-Dirac
representation of the Dirac matrices $\hat {\bm \alpha}$ and $\hat
{\beta}$, in which
\begin{equation}\label{a_b}
    \hat {\bm \alpha}=
\begin{pmatrix}{0}&{\hat {\bm \sigma}} \\
{\hat {\bm \sigma}}& {0}
\end{pmatrix}=\gamma_0{\bm \gamma}, \ \ \
\hat {\beta}=\begin{pmatrix}{1}&{0} \\
{0}& {-1}
\end{pmatrix}=\gamma_0, \ \ {\hat {\bm \Sigma}}=
\begin{pmatrix}{\hat {\bm \sigma}}&{0} \\
{0}&{\hat {\bm \sigma}}
\end{pmatrix},
\end{equation}
where ${\hat {\bm \sigma}}=({ \sigma}_{1},{ \sigma}_{2},{
\sigma}_{3})$ and  $\sigma$ denotes the Pauli matrixes.

Now let us consider the case of a neutrino moving in matter
without any electromagnetic field in the background. The quantum
equation for the neutrino wave function can be obtained from
(\ref{D_P}) with application of the substitution (\ref{sub}) which
now becomes

\begin{equation}\label{sub_1}
  F_{\mu \nu}\rightarrow G_{\mu \nu}.
\end{equation}
Thus, we get the quantum equation for the neutrino wave function
in the presence of the background matter in the form
\begin{equation}\label{D_P_matter}
\Big( i\gamma^{\mu}\partial_{\mu} - m -\frac{\mu}{2}\sigma ^{\mu
\nu}G_{\mu \nu}\Big)\Psi(x)=0,
\end{equation}
that can be regarded as the generalized Dirac-Pauli equation. The
generalization of the neutrino quantum equation for the case when
an electromagnetic field is present, in addition to the background
matter, is obvious.

The detailed discussion on the evaluation of the tensor $G_{\mu
\nu}$ is given in \cite{EgoLobStu00LobStu01DvoStuJHEP02StuPAN04}.
We consider here, for simplicity,  the case of the unpolarized
matter composed of the only one type of fermions  of a constant
density. For a background of only electrons we  get
\begin{equation}\label{G_1}
G^{\mu \nu}= \gamma \rho^{(1)} n
\begin{pmatrix}{0}&{0}& {0}&{0} \\
{0}& {0}& {-\beta_{3}}&{\beta_{2}} \\
{0}&{\beta_{3}}& {0}&{-\beta_{1}} \\
{0}&{-\beta_{2}}& {\beta_{1}}& {0}
\end{pmatrix}, \gamma=(1-{\bm\beta}^{2})^{-1/2},
\rho^{(1)}=\frac{\tilde{G}_{F}}{2\sqrt{2}\mu},
\tilde{G}_{F}=G_F(1+4\sin^{2}\theta_{W}),
\end{equation}
where ${\bm \beta}=(\beta_{1},\beta_{2},\beta_{3})$ is the
neutrino three-dimensional speed, $n$ denotes the number density
of the background electrons. From (\ref{G_1}) and the two
equations, (\ref{D_P}) and (\ref{D_P_matter}), it is possible to
see that the term $\gamma \rho^{(1)}n{\bm \beta}$ in
Eq.(\ref{D_P_matter}) plays the role of the magnetic field $\bf B$
in Eq.(\ref{D_P}). Therefore, the Hamiltonian form of
(\ref{D_P_matter}) is
\begin{equation}\label{H_G_form}
i\frac{\partial}{\partial t}\Psi({\bf r},t)=\hat H_{G}\Psi({\bf
r},t),
\end{equation}
where
\begin{equation}\label{H_G}
  \hat H_{G}=\hat {\bm{\alpha}} {\bf p} + \hat {\beta}m + \hat V_G,
\end{equation}
and
\begin{equation}\label{V_G}
\hat V_G= -\mu  \frac{\rho^{(1)} n}{m}
  \hat {\beta}{\bf \Sigma}{\bf p},
\end{equation}
here $\bf p$ is the neutrino momentum. From (\ref{V_G}) it is just
straightforward that the potential energy in matter depends on the
neutrino helicity.

The form of the Hamiltonian (\ref{H_G}) ensures that the operators
of the momentum, $\hat {\bf p}$, and helicity, ${\bf \Sigma} {\bf
p}/p$, are integrals of motion. That is why for the stationary
states we can write
\begin{equation}\label{stat_states}
\Psi({\bf r},t)=e^{-i( Et-{\bf p}{\bf r})}u({\bf p},E),
\ \ \ u({\bf p},E)=\begin{pmatrix}{\varphi}\\
{\chi}
\end{pmatrix},
\end{equation}
where $u({\bf p},E)$ is independent on the coordinates and time
and can be expressed in terms of the two-component spinors
$\varphi$ and $\chi$. Substituting (\ref{stat_states}) into
Eq.(\ref{H_G_form}), we get the two equations
\begin{equation}\label{H_u}
    ({\bm \sigma}{\bf p})\chi-\big(E-m+\alpha({\bm \sigma}{\bf
    p})\big)\varphi=0, \ \ \
({\bm \sigma}{\bf p})\varphi-\big(E-m+\alpha({\bm \sigma}{\bf
    p})\big)\chi=0.
\end{equation}
Suppose that $\varphi$ and $\chi$ satisfy the following equations,
\begin{equation}\label{helicity}
  ({\bm \sigma}{\bf p})\varphi=sp\varphi, \ \
  ({\bm \sigma}{\bf p})\chi=sp\chi,
\end{equation}
where $s=\pm 1$ specify the two neutrino helicity states. Upon the
condition that the set of Eqs.(\ref{H_u}) has a non-trivial
solution, we arrive to the energy spectrum of a neutrino moving in
the background matter:
\begin{equation}\label{Energy}
  E={\sqrt{{\bf p}^{2}(1+\alpha^{2})+m^2 -2\alpha m p s}},
\ \  \ \   \alpha=\frac{\mu \rho^{(1)}}{m}n=
  \frac{1}{2\sqrt{2}}{\tilde G}_{F}\frac{n}{m}.
\end{equation}
 It is important that the the neutrino energy in the
background matter depends on the state of the neutrino
longitudinal polarization (helicity), i.e. the left-handed and
right-handed neutrinos with equal momentum have different
energies.

The obtained expression  (\ref{Energy}) for the neutrino energy
can be transformed to the form
\begin{equation}\label{Energy_1}
E=\sqrt{{\bf p}^{2}+m^2\Big(1-s\frac{\alpha p}{m}\Big)^{2}}.
\end{equation}
It is easy to see that the energy spectrum of a neutrino in
vacuum, which is derived on the basis of the Dirac equation, is
modified in the presence of matter by the formal shift of the
neutrino mass
\begin{equation}\label{substitution}
  m\rightarrow m\Big(1-s\frac{\alpha p}{m}\Big).
\end{equation}

The procedure, similar to one used for the derivation of the
solution of the Dirac equation in vacuum, can be adopted for the
case of the neutrino moving in matter. We apply this procedure to
the equation (\ref{H_G_form}) and arrive to the final form of the
wave function of a neutrino moving in the background matter:
\begin{equation}\label{wave_function}
\Psi_{{\bf p},s}({\bf r},t)=\frac{e^{-i( Et-{\bf p}{\bf
r})}}{2L^{\frac{3}{2}}}
\begin{pmatrix}{\sqrt{1+ \frac{m-s\alpha p}{E}}}
\ \sqrt{1+s\frac{p_{3}}{p}}
\\
{s \sqrt{1+ \frac{m-s\alpha p}{E}}} \ \sqrt{1-s\frac{p_{3}}{p}}\ \
e^{i\delta}
\\
{  s\sqrt{1- \frac{m-s\alpha p}{E}}} \ \sqrt{1+s\frac{p_{3}}{p}}
\\
{\sqrt{1- \frac{m-s\alpha p}{E}}} \ \ \sqrt{1-s\frac{p_{3}}{p}}\
e^{i\delta}
\end{pmatrix} ,
\end{equation}
where $L$ is the normalization length and $ \delta=\arctan
p_{y}/p_{x}$. In the limit of vanishing density of matter, when
$\alpha\rightarrow 0$, the wave function of
Eq.(\ref{wave_function}) transforms to the vacuum solution of the
Dirac equation.

The proposed new quantum equation (\ref{D_P_matter}) for a
neutrino moving in the background matter and the obtained exact
solutions (\ref{wave_function}) establish a basis for a new method
in the study of different processes with participation of
neutrinos in the presence of matter. As an example, we should like
to use the new method in the study of the spin light of neutrino
($SL\nu$) in matter and to develop the {\it quantum} theory of
this effect. Within the quantum approach, the corresponding
Feynman diagram of the $SL\nu$ in matter is the standard
one-photon emission diagram with the initial and final neutrino
states described by the "broad lines" that account for the
neutrino interaction with matter. From the usual neutrino magnetic
moment interaction, it follows that the amplitude of the
transition from the neutrino initial state $\psi_{i}$ to the final
state $\psi_{f}$, accompanied by the emission of a photon with a
momentum $k^{\mu}=(\omega,{\bf k})$ and a polarization ${\bf
e}^{*}$, can be written in the form
\begin{equation}\label{amplitude}
  S_{f i}=-\mu \sqrt{4\pi}\int d^{4} x {\bar \psi}_{f}(x)
  ({\hat {\bm \Gamma}}{\bf e}^{*})\frac{e^{ikx}}{\sqrt{2\omega L^{3}}}
   \psi_{i}(x),
\end{equation}
where $\psi_{i}$ and $\psi_{f}$ are the corresponding exact
solutions of the equation (\ref{D_P_matter}) given by
(\ref{wave_function}), and
\begin{equation}\label{Gamma}
  \hat {\bm \Gamma}=i\omega\big\{\big[{\bm \Sigma} \times
  {\bm \varkappa}\big]+i\gamma^{5}{\bm \Sigma}\big\}.
\end{equation}
Here ${\bm \varkappa}=\frac{\bf k}{\omega}$ is the unit vector
pointing in the direction of the emitted photon propagation.

The integration in (\ref{amplitude}) with respect to time yields
\begin{equation}\label{amplitude}
  S_{f i}=-\mu {\sqrt {\frac {2\pi}{\omega L^{3}}}}
  2\pi\delta(E_{f}-E_{i}+\omega)
  \int d^{3} x {\bar \psi}_{f}({\bf r})({\hat {\bf \Gamma}}{\bf e}^{*})
  e^{i{\bf k}{\bf r}}
   \psi_{i}({\bf r}),
\end{equation}
where the delta-function stands for the energy conservation.
Performing the integrations over the spatial co-ordinates, we can
recover the delta-functions for the three components of the
momentum. Finally, we get the law of energy-momentum conservation
for the considered process,
\begin{equation}\label{e_m_con}
    E_{i}=E_{f}+\omega, \ \
    {\bf p}_{i}={\bf p}_{f}+{\bm \varkappa}.
\end{equation}

Let us suppose that the weak interaction of the neutrino with the
electrons of the background is indeed weak. Here we should like to
note that Eq.(\ref{D_P_matter}) is derived under the assumption
that the matter term is small. This condition is similar to the
condition of smallness of the electromagnetic term in the
Dirac-Pauli equation (\ref{D_P}) in the electrodynamics. In this
case, we can expand the energy (\ref{Energy_1}) over $\alpha\frac
{pm}{E_{0}^{2}} \ll 1$ and in the liner approximation get
\begin{equation}\label{Energy_2}
  E\approx E_{0}-sm\alpha \frac{p}{E_{0}},
\end{equation}
where $E_0=\sqrt{p^2 +m^2}$. Then from the law of the energy
conservation (\ref{e_m_con}) we get for the energy of the emitted
photon
\begin{equation}\label{omega}
  \omega=E_{{i}_{0}}-{E}_{{f}_{0}}+\Delta,
  \ \Delta=\alpha m \frac{p}{E_0}(s_{f}-s_{i}),
\end{equation}
where the indexes $i$ and $f$ label the corresponding quantities
of the neutrino initial and final states. From Eqs.(\ref{omega})
and the law of the momentum conservation, in the linear
approximation over $\alpha$, we obtain
\begin{equation}\label{omega_1}
    \omega=(s_{f}-s_{i})\alpha m \frac {\beta}{1-\beta \cos
    \theta},
\end{equation}
where $\theta$ is the angle between ${\bm \varkappa}$ and the
direction of the neutrino speed ${\bm \beta}$.

From the above consideration it follows that the only possibility
for the $SL\nu$ to appear is provided in the case when the
neutrino initial and final states are characterized by $s_{i}=-1$
and $s_{f}=+1$, respectively. Thus we conclude, on the basis of
the quantum treatment of the $SL\nu$ in matter, that in this
process the left-handed neutrino is converted to the right-handed
neutrino (see also \cite{LobStuPLB03}) and the emitted photon
energy is given by
\begin{equation}\label{omega_12}
    \omega=\frac{1}{\sqrt{2}}{\tilde G}_{F}n \frac {\beta}{1-\beta \cos
    \theta}.
\end{equation}
Note that the photon energy depends on the angle $\theta$ and also
on the value of the neutrino speed $\beta$. In the case of
$\beta\approx 1$ and $\theta\rightarrow 0$ we confirm the
estimation for  the emitted photon energy given in
\cite{LobStuPLB03}.

Now let us derive the $SL\nu$ rate and radiation power using the
quantum theory. In the case of a neutrino moving along the
OZ-axes, the solution (\ref{wave_function}) for the states with
$s=-1$ and $s=+1$ can be rewritten in the form,
\begin{equation}\label{wave_function_min}
\Psi_{{\bf p},s=-1}({\bf r},t)=\frac{e^{-i(Et-{\bf p}{\bf
r})}}{\sqrt{2}L^{\frac{3}{2}}}
\begin{pmatrix}{0}
\\
{- \sqrt{1+ \frac{m+\alpha p}{E}}} \
\\
{ 0}
\\
{\sqrt{1- \frac{m+\alpha p}{E}}}
\end{pmatrix} ,
\end{equation}
and
\begin{equation}\label{wave_function_min}
\Psi_{{\bf p},s=+1}({\bf r},t)=\frac{e^{-i(Et-{\bf p}{\bf
r})}}{\sqrt{2}L^{\frac{3}{2}}}
\begin{pmatrix}{ \sqrt{1+ \frac{m-\alpha p}{E}}}
\\
{0} \
\\
{ \sqrt{1- \frac{m-\alpha p}{E}}}
\\
{0}
\end{pmatrix}.
\end{equation}
We now put these wave functions into Eq.(\ref{amplitude}) and
calculate the spin light transition rate in the linear
approximation over the parameter $\alpha\frac {pm}{E_{0}^{2}}$.
Finally, for the rate we get
\begin{equation}\label{rate}
  \Gamma_{SL}=8
  \mu^{5}(n\rho^{(1)}\beta)^{3}\int
  \limits_{}^{}\frac{S\sin \theta}{(1-\beta \cos \theta)^{4}}
   d\theta,
\end{equation}
where
\begin{equation}\label{S}
S=(\cos \theta - \beta)^{2}+
  (1-\beta \cos \theta)^{2}.
\end{equation}
The corresponding expression for the radiation power is
\begin{equation}\label{power}
  I_{SL}=16\mu^{6}(n\rho^{(1)}\beta)^{4}
  \int\limits_{}^{}\frac{S\sin \theta}{(1-\beta \cos \theta)^{5}}
   d\theta.
\end{equation}
Performing the integrations in Eq.(\ref{rate}) over the angle
$\theta$, we obtain for the rate
\begin{equation}\label{rate_1}
  \Gamma_{SL}=\frac{2\sqrt{2}}{3}\mu^{2}{{\tilde G}_{F}}^{3}
  n^{3}\beta^{3}\gamma^{2}.
\end{equation}
This result exceeds the value of the neutrino spin light rate
derived in \cite{LobStuPLB03} by a factor of two because here the
neutrinos in the initial state  are totally left-handed polarized,
whereas in \cite{LobStuPLB03} the case of initially unpolarized
neutrinos (i.e., an equal mixture of the left- and right-handed
neutrinos) is considered. From Eq.(\ref{power}) we get for the
total radiation power,
\begin{equation}\label{power_1}
  I_{SL}=\frac{2}{3}\mu^{2}{{\tilde G}_{F}}^{4}
  n^{4}\beta^{4}\gamma^{4}.
\end{equation}

Using Eq.(\ref{amplitude}) we can also derive the $SL\nu$ rate and
total power in matter accounting for the photon polarization. If
 $\bf j$ is the unit vector pointing in the direction of the neutrino
propagation, then we can introduce the to vectors
\begin{equation}\label{e_12}
  {\bf e}_1= \frac{[{\bm \varkappa}\times {\bf j}]}
  {\sqrt{1-({\bm \varkappa}{\bf j})^{2}}}, \ \
  {\bf e}_2= \frac{{\bm \varkappa}({\bm \varkappa}{\bf j})-{\bf j}}
  {\sqrt{1-({\bm \varkappa}{\bf j})^{2}}},
\end{equation}
which specify the two different linear polarizations of the
emitted photon.  For these vectors it is easy to get
\begin{equation}\label{e_12a}
{\bf e}_1=\{\sin \phi, -\cos \phi\}, \ \ {\bf e}_2=\{\cos \phi
\cos \theta, \sin \phi \cos \theta, -\sin \theta \}.
\end{equation}
Note that the vector ${\bf e}_1$ is orthogonal to ${\bf j}$.
Decomposing the neutrino transition amplitude (\ref{amplitude}) in
contributions from the two linearly polarized photons, one can
obtain the power of the process with radiation of the polarized
photons. For the two linear polarizations determined by the
vectors ${\bf e}_1$ and ${\bf e}_2$, we get
\begin{equation}\label{power_pol}
    I_{SL}^{(1),(2)}=16 \mu^{6}(n\rho^{(1)}\beta)^{4}
  \int\limits_{}^{}\frac{\sin \theta}{(1-\beta \cos \theta)^{5}}
 \begin{pmatrix}{S^{(1)}}
\\
{S^{(2)}} \
\end{pmatrix}
d\theta,
\end{equation}
where
\begin{equation}\label{S_12}
S^{(1)}=(\cos \theta - \beta)^{2}, \ \ S^{(2)}=(1-\beta \cos
\theta)^{2}.
\end{equation}

Finally, performing the integration over the angle $\theta$ we get
the power of the radiation of the linearly polarized photons
\begin{equation}\label{power_1}
  \begin{pmatrix}{I_{SL}^{(1)}}
  \\
{I_{SL}^{(2)}}
\end{pmatrix}
  =\begin{pmatrix}{\frac{1}{3}}
  \\
{1}
\end{pmatrix}\frac{1}{2}\mu^{2}
{{\tilde G}_{F}}^{4}n^{4}\beta^{4}\gamma^{4}.
\end{equation}

It is also possible to decompose the radiation power for the
circular polarized photons. We introduce the two unit vectors for
description of the photons with the two opposite circular
polarizations,
\begin{equation}\label{circ_pol}
  {\bf e}_{l}=\frac{1}{\sqrt 2}({\bf e}_{1}\pm i{\bf e}_{2})
\end{equation}
where $l=\pm 1$ correspond to the right and left photon circular
polarizations, respectively. Then for the power of the radiation
of the circular-polarized photons we get
\begin{equation}\label{power_pol_circ}
    I_{SL}^{(l)}=16 \mu^{6}(n\rho^{(1)}\beta)^{4}
  \int\limits_{}^{}\frac{\sin \theta}{(1-\beta \cos \theta)^{5}}
 {S^{(l)}}
d\theta,
\end{equation}
where
\begin{equation}\label{S_l}
S^{(l)}=\frac{1}{2}(S^{(1)}+S^{(2)})-l\sqrt{S^{(1)}S^{(2)}}.
\end{equation}
Integration over the angle $\theta$ in (\ref{power_pol_circ})
yields
\begin{equation}\label{power_l}
  {I_{SL}^{(l)}}
    =\frac{1}{3}\mu^{2}
{{\tilde G}_{F}}^{4}n^{4}\beta^{4}\gamma^{4}
    \Big(1-\frac{1}{2}l\beta\Big).
\end{equation}

Note that information on the  photons polarization may be
important for the experimental observation of the $SL\nu$ from
different astrophysical and cosmology media.

In conclusion, we have shown how the Dirac-Pauli equation for a
massive neutrino in an external electromagnetic field can be
modified in order to account for the effect of the neutrino-matter
interaction. On the basis of the new equation the quantum
treatment of a neutrino moving in the presence of the background
matter has been realized. In the limit of a constant density of
the matter, we have obtained the exact solutions of this new
equation for different neutrino helicity states and also
determined the neutrino energy spectrum, which depends on the
helicity. Then we have developed the quantum theory of the $SL\nu$
in the matter and calculated its rate and power accounting for the
emitted photons polarization. We  have also confirmed, within the
solid base of the developed quantum approach, the existence of the
neutrino-self polarization effect \cite{LobStuPLB03,LobStuPLB04}
in the process of the spin light radiation of a neutrino moving in
the background matter. The $SL\nu$ radiation and the corresponding
neutrino self-polarization effect, due to the significant
dependence on the matter density, are expected to be important in
different astrophysical dense media and in the early Universe.

The authors are thankful to Alexander Grigoriev for useful
discussions.

\end{document}